\documentclass[a4paper,11pt]{article}
\pdfoutput=1
\usepackage{jheppub}

\usepackage{amssymb,amsmath}
\usepackage[normalem]{ulem}
\usepackage[utf8x]{inputenc}
\usepackage{slashed}
\usepackage{bm}
\usepackage{graphicx}

\usepackage{csquotes} 
\usepackage{comment}
\usepackage{mathrsfs}
\usepackage{float}

\usepackage{ascmac}

\def\lsim{\mathrel{\rlap{\lower4pt\hbox{\hskip1pt$\sim$}}
    \raise1pt\hbox{$<$}}}         
\def\gsim{\mathrel{\rlap{\lower4pt\hbox{\hskip1pt$\sim$}}
    \raise1pt\hbox{$>$}}}         

\numberwithin{equation}{section}

\preprint{
\begin{minipage}{5cm}
\small
\flushright
KEK-TH-2542\\KYUSHU-HET-267\\KOBE-TH-23-04
\end{minipage}}

\title{Upper bound on the Atiyah-Singer index from tadpole cancellation}

\author{Keiya Ishiguro$^{1}$,} 
\author{Takafumi Kai$^{2}$,} 
\author{Satsuki Nishimura$^{2}$,} 
\author{Hajime Otsuka$^{2}$, and}  
\author{Maki Takeuchi$^{3}$} 

\affiliation{
$^1$ Graduate University for Advanced Studies (Sokendai), 1-1 Oho, Tsukuba, Ibaraki 305-0801, Japan\\
$^2$ Department of Physics, Kyushu University, 744 Motooka, Nishi-ku, Fukuoka 819-0395, Japan\\
$^3$ Department of Physics, Kobe University, Kobe 675-8501, Japan\\
}
\emailAdd{kai.takafumi.606@s.kyushu-u.ac.jp}
\emailAdd{ishigu@post.kek.jp}
\emailAdd{nishimura.satsuki@phys.kyushu-u.ac.jp}
\emailAdd{191s107s@stu.kobe-u.ac.jp}
\emailAdd{otsuka.hajime@phys.kyushu-u.ac.jp}

\abstract{
We propose an upper bound on the Atiyah-Singer index in the effective action of string theory. 
For $E_8\times E_8^\prime$ and $SO(32)$ heterotic string theories on smooth Calabi-Yau threefolds with line bundles, 
we find that the tadpole cancellation and supersymmetry conditions lead to an upper bound on the generation number of quarks and leptons as well as Higgs doublets. 
By taking into account the observed value of four-dimensional gauge couplings and the supergravity approximation, we explicitly 
evaluate the bound on favorable complete intersection Calabi-Yau threefolds. 
The bound can be extended to Calabi-Yau threefolds in the Kreuzer-Skarke database. 
We also put the upper bound on the Atiyah-Singer index in Type IIB/F-theory compactifications.
}
\makeatletter
\gdef\@fpheader{}
\makeatother

\begin{document}

\maketitle

\section{Introduction}
\label{sec:intro}

The origin of fermion generations is one of the unsolved problems in the Standard Model (SM). 
Higher-dimensional theories have been utilized to understand the generation structure of chiral fermions. 
Indeed, background fluxes and curvatures in the extra-dimensional space will lead to degenerate chiral fermions in the four-dimensional (4D) low-energy effective action through the Kaluza-Klein compactifications \cite{Candelas:1985en}. 
The degeneracy of chiral zero modes will be counted by the Atiyah-Singer index theorem, but 
an arbitrary value of background fluxes is allowed in general. 
Thus, there is no guiding principle to fix the chiral index, and 
it will be a challenging problem to answer the generation problem of quarks and leptons.

\medskip

In string theory, brane charges induced by the background fluxes and curvatures 
should be canceled in compact spaces, the so-called tadpole cancellation conditions. 
It suggests that the possibility of background fluxes is finite through the cancellation of brane charges, 
and the Atiyah-Singer index determining the number of chiral zero modes is bounded by some topological quantities. 
There was an attempt to put the bound on the chiral index in the context of heterotic $E_8\times E_8^\prime$ string theory 
on Calabi-Yau (CY) threefolds with line bundles \cite{Buchbinder:2013dna,Constantin:2015bea,GrootNibbelink:2015dvi} under the assumption that the effective action is described by the 4D ${\cal N}=1$ supergravity 
and the four-dimensional coupling constants are finite. 
This bound was derived by the requirement of the supersymmetry and anomaly cancellation. 
It motivates us to explore the maximal number of quarks and leptons in other corners of string theories. 

\medskip

The purpose of this paper is to derive a rigorous bound on the Atiyah-Singer index in the effective action of string theory. 
We focus on three classes of string effective action: (i) $E_8\times E_8^\prime$ heterotic line bundle models incorporating the 4D $SU(5)$ grand unified theory (GUT), (ii) $SO(32)$ heterotic line bundle models incorporating the 4D Pati-Salam model or SM-like model, (iii) Type IIB magnetized D7-brane models lift to the F-theory vacua on compact CY fourfolds. 
In these models, Abelian fluxes lead to degenerate chiral zero modes counted by the Atiyah-Singer index theorem, 
but the degeneracy will be correlated with the brane charges due to the tadpole cancellation conditions. 
Thus, the tadpole charge puts a severe bound on the chiral index. 
Our findings about the bound on the index of chiral matters are summarized as follows:

\begin{itemize}

\item For $E_8\times E_8^\prime$ heterotic line bundle models, we find that the index of quarks and leptons is bounded from above by 
the tadpole cancellation condition:
\begin{align}
        |N_{\rm gen}| <  \alpha |m_{\rm max}| \|c_{2}(T{\cal M})\|,
        \label{eq:intro_bound}
\end{align}
where $\alpha$ is the ${\cal O}(1)$ constant, $m_{\rm max}$ is the largest $U(1)$ flux quanta and $\|c_{2}(T{\cal M})\|$ denotes the Euclidean norm of the second Chern number of CY threefolds. 
Note that $m_{\rm max}$ is also bounded from above by CY topological data through the supersymmetry, K-theory and tadpole cancellation conditions. 
When we move to $SO(32)$ heterotic line bundle models, we arrive at a similar bound on the index of Higgs doublets and vector-like quarks under the SM gauge group.

\item We explicitly check the upper bound on the chiral index on favorable complete intersection CY threefolds \cite{Candelas:1987kf,Candelas:1987du}, 
taking into account the finite value of 4D coupling constants. 
The analysis can be extended to the CY threefolds in the Kreuzer-Skarke database \cite{Kreuzer:2000xy}.

\item We explore the bound of the chiral index in the context of Type IIB/F-theory compactifications. 
Since the index of chiral zero modes arising from the intersection between two stacks of magnetized D7-branes is determined by 
world-volume fluxes on D7-branes, the chiral index is bounded by the brane charges through the tadpole cancellation 
condition of D3-branes.

\end{itemize}

This paper is organized as follows. 
First, we revisit the $E_8\times E_8^\prime$ heterotic line bundle models. 
After summarizing the consistency conditions in the heterotic model building, 
we derive the upper bound on the chiral index in $E_8\times E_8^\prime$ heterotic 
line bundle models in Sec. \ref{sec:E8}. 
For $SO(32)$ heterotic line bundle models, we consider two scenarios: (i) 4D Pati-Salam 
models in Sec. \ref{sec:pati-salam} and (ii) 4D SM-like models realized by the so-called 
hypercharge flux breaking. 
In addition to the chiral index of quarks/leptons, the index of 
vector-like quarks and Higgs doublets is also bounded from above by geometric quantities of CY 
threefolds. 
In Sec. \ref{sec:typeII}, we analyze the chiral index on magnetized D7-branes in Type IIB/F-theory compactifications. 
Sec. \ref{sec:con} is devoted to the conclusions.

\section{$E_8\times E_8^\prime$ heterotic string theory}
\label{sec:E8}

We start from $E_8\times E_8^\prime$ heterotic string theory on CY threefolds with 
line bundles.\footnote{For more details, see, Refs. \cite{Blumenhagen:2005ga,Blumenhagen:2005pm}.} 
The bosonic part of low-energy effective action up to the order of $\alpha^\prime$ is described by
\begin{align}
S_{\rm bos}&=\frac{1}{2\kappa_{10}^2}\int_{M^{(10)}} 
e^{-2\phi_{10}} \left[ R+4d\phi_{10} \wedge  \ast 
d\phi_{10} -\frac{1}{2}H\wedge \ast H \right] 
\nonumber\\
&-\frac{1}{2g_{10}^2} \int_{M^{(10)}} e^{-2\phi_{10}} 
{\rm tr}(F\wedge \ast F)
-\sum_s N_s T_5 \int_{M^{(10)}} B^{(6)} \wedge \delta(\gamma_s), 
\label{eq:heterob}
\end{align}
with
\begin{align}
 H=dB^{(2)} -\frac{\alpha^{'}}{4}(w_{\rm YM} -w_{L}),   
\end{align}
consisting of the Kalb-Ramond two-form $B^{(2)}$ (whose hodge dual is $B^{(6)}$) and the gauge and gravitational Chern-Simons three-forms, $w_{\rm YM}$ and $w_{L}$. 
Here, $\phi_{10}$ denotes the dilaton, and the 10D gravitational coupling $\kappa_{10}$ and gauge coupling $g_{10}$ are normalized as $2\kappa_{10}^2=(2\pi)^7(\alpha')^4$ 
and $g_{10}^2=2(2\pi)^7(\alpha')^3$, respectively. 
The trace of 10D gauge field strength $F$ is taken in the fundamental representation of $E_8$. 
Furthermore, we introduce $N_s$ stacks of heterotic five-branes wrapping the holomorphic two-cycles $\gamma_s$ whose Poincar\'{e} dual four-form is represented by $\delta (\gamma_s)$, and the tension of five-branes is given by $T_5=((2\pi)^5(\alpha')^3)^{-1}$. 

\medskip

To realize the semi-realistic spectra, we introduce the internal gauge bundle consisting of 
multiple line bundles $l_a$, each with a structure group $U(1) \subset E_8$, that is,
\begin{align}
    W = \bigoplus_{a=1}^n l_a.
\end{align}
These line bundles play a role not only in breaking the $E_8$ gauge group to $G \times \Pi_a U(1)_a$ but also 
in generating the chiral fermions from the adjoint representation of $E_8$:
\begin{align}
    248 \rightarrow \bigoplus_p (R_p, C_p),
\end{align}
under $G$ and $\Pi_a U(1)_a$, respectively. 
Indeed, the index of chiral massless fermions (chiral superfields if the supersymmetry is preserved) in the representation $R_p$ is counted by the Atiyah-Singer index:
\begin{align}
    \chi({\cal M}, {\cal C}_p)
    = \int_{\cal M} \left( {\rm ch}_3({\cal C}_p) + \frac{1}{12}c_2(T{\cal M})c_1({\cal C}_p)\right),
\end{align}
where ${\rm ch}_3({\cal C}_p)$ and $c_2(T{\cal M})$ denote the third Chern character of the internal bundle of each $C_p$ 
and the second Chern class of the tangent bundle of CY threefolds ${\cal M}$, respectively. 
However, the line bundles should satisfy four consistency conditions, as will be discussed below.

\begin{enumerate}
    \item Tadpole cancellation condition

    From the Bianchi identity of the Kalb-Ramond field $B^{(6)}$:
    \begin{align}
d(e^{2\phi_{10}}\ast dB^{(6)}) =-\frac{\alpha^{'}}{4} 
\left({\rm tr}\bar{F}^2 -{\rm tr}{\bar R}^2 -
4(2\pi)^2 \sum_s N_s \delta(\gamma_s) \right)
\end{align}
with $\Bar{F}$ and $\Bar{R}$ being the internal background field strengths, 
the line bundles satisfy the following inequality in cohomology:
\begin{align}
{\rm ch}_2(W)+c_2(T{\cal M})= \sum_s N_s \delta(\gamma_s) \geq [0].
\label{eq:tad}
\end{align}
Here and in what follows, the trace of $R$ is taken in the fundamental representation of $SO(1,9)$. 

    \item K-theory condition

    The first Chern class of the total bundle $W$ is constrained as
    \begin{align}
        c_1(W) = \sum_{a=1}^n n_a c_1(l_a) \in H^2({\cal M}, 2\mathbb{Z})
        \label{eq:K-theory}
    \end{align}
    to cancel the anomalies in the two-dimensional non-linear sigma model \cite{Witten:1985mj,Freed:1986zx}. 
    This condition will be regarded as the K-theory condition in Type I string theory \cite{Witten:1998cd,Uranga:2000xp}. 
    In what follows, $n_a$ is normalized as one in the analysis of $E_8\times E_8^\prime$ heterotic 
    string theory, but it depends on the embedding of line bundles, as shown in Sec. \ref{sec:SO(32)} for $SO(32)$ 
    heterotic string theory. 
    Throughout this paper, we focus on the case with $c_1(W)=0$, corresponding to a 
    structure group $S(U(1)^n)$.

     \item Supersymmetry condition

     The internal gauge field strength $\Bar{F}$ has to be holomorphic, that is,
     \begin{align}
\bar{F}_{ij}=\bar{F}_{\bar{i}\bar{j}}=0,\qquad
g^{i\bar{j}}\bar{F}_{i\bar{j}}=0,
\end{align}
     due to the supersymmetry condition. The so-called Hermitian Yang-Mills equation can 
     be solved when
    \begin{align}
\mu(l_a)=\frac{1}{2l_s^4}\int_{\cal M} J\wedge J\wedge c_1(l_a) =0\qquad \forall a,
\label{eq:Dterm}
\end{align}
    at the leading order. 
    Here, $l_s$ denotes the string length, and $J$ denotes the K\"ahler form of CY threefolds. 

    \item Bound on the 4D gauge coupling

     From the dimensional reduction of Einstein-Hilbert term on a six-dimensional internal manifold, the 4D Planck mass $M_{\rm Pl}$ is extracted as
     \begin{align}
         M_{\rm Pl}^2 = e^{-2\langle \phi_{10}\rangle} \frac{{\rm Vol}({\cal M})}{\kappa_{10}^2} 
         = 4\pi \frac{{\cal V}}{g_s^2 l_s^2},
     \end{align}
     with $g_s = e^{\langle \phi_{10}\rangle}$ being the string coupling. 
     Here, the CY volume is measured in units of string length, ${\rm Vol}({\cal M})={\cal V}\,l_s^6$. 
     Since the 4D gauge coupling is derived from the dimensional reduction of 10D gauge field strength
     \begin{align}
         \alpha^{-1} = \frac{4\pi}{g_4^2} = e^{-2\langle \phi_{10}\rangle}\frac{4\pi {\rm Vol}({\cal M})}{g_{10}^2} = g_s^{-2} {\cal V},
     \end{align}
     the internal volume is bounded from above by
     \begin{align}
         {\cal V}=g_s^2 \alpha^{-1} \lesssim 25,
     \end{align}
     where we adopt the 4D gauge coupling at the GUT scale, $\alpha^{-1}(M_{\rm GUT}) \simeq 25$ with $M_{\rm GUT} \simeq 2\times 10^{16}$\,GeV. 
     We recall that the CY volume is described by
     \begin{align}
         {\cal V}= \frac{1}{6}\int_{\cal M} J \wedge J \wedge J = \frac{1}{6}\sum_{i,j,k} d_{ijk}t^it^jt^k,
         \label{eq:CYvolume_bound}
     \end{align}
     with $J=\sum_{i=1}^{h^{1,1}}t^iw_i$ in a basis $w_i$ of $H^{1,1}({\cal M},\mathbb{R})$. 
     Here, $t^i$ and $d_{ijk}$ denote the K\"ahler moduli and the triple intersection numbers of CY manifolds, that is, $d_{ijk}=\int_{\cal M}w_i \wedge w_j\wedge w_k$, respectively. 
     Note that the K\"ahler moduli should reside in $t^i> 1$ in units of $l_s$, otherwise 
     the supergravity description is not valid. 
     Thus, the bound on the CY volume (\ref{eq:CYvolume_bound}) restricts the sum of intersection number as follows:
     \begin{align}
         \sum_{i,j,k}d_{ijk} < \sum_{i,j,k}d_{ijk}t^it^jt^k = 6{\cal V} \leq 150.
         \label{eq:Volume_bound}
     \end{align}
Here, we assume non-negative triple intersection numbers as in the favorable complete intersection Calabi-Yaus.

\end{enumerate}

In the following, we restrict the index of chiral zero modes by using the above consistency 
conditions in $E_8\times E_8^\prime$ heterotic string theory on CY threefolds with 
line bundles. 
In particular, we focus on the $SU(5) \times S(U(1)^5)$ GUT as a branch of $E_8$.\footnote{See Refs. \cite{Anderson:2011ns,Anderson:2012yf} for a detailed analysis.} 
The net number of chiral zero modes is then described by
\begin{align}
N_{\rm gen} = - {\rm ind}(W) = \frac{1}{2} \int_{\cal M} c_3(W) = -\sum_{i,j,k}\sum_{a=1}^n\frac{d_{ijk}}{6} m_a^i m_a^j m_a^k,    
\label{eq:Index_SU(5)}
\end{align}
where the first Chern classes are of the form
\begin{align}
    c_1(l_a) = \sum_{i=1}^{h^{1,1}} m_a^i w_i, 
\end{align}
in the basis $w_i$ of $H^{1,1}({\cal M},\mathbb{R})$. 
Here, $m_a^i$ denotes the quantized integer, and the number of $U(1)$ is taken as an arbitrary integer $n$, although $n=5$ in the case of $SU(5) \times S(U(1)^5)$ GUT. 
Following Ref. \cite{Constantin:2015bea}, we present the bound of line bundle vector $\bm{m_a}$. 
Since the moduli metric:
\begin{align}
    G_{ij} &= \frac{1}{2(2\pi l_s)^6{\cal V}} \int_{\rm CY} J_i \wedge \ast J_j 
    = -\frac{1}{2}\partial_i \partial_j \ln {\cal V}
    =-3\left( \frac{\kappa_{ij}}{\kappa} - \frac{3\kappa_i\kappa_j}{2\kappa^2}\right),
\end{align}
with
\begin{align}
    \kappa= 6{\cal V} = \sum_{i,j,k}d_{ijk}t^it^jt^k,\qquad
    \kappa_i = \sum_{j,k}d_{ijk}t^jt^k,\qquad
    \kappa_{ij} = \sum_{k} d_{ijk}t^k
\end{align}
is positive definite in the interior of the K\"ahler cone, one can derive the following inequality:
\begin{align}
    0 &< \sum_{a,i} m_a^i G_{ij} m_a^j = -\frac{3}{\kappa} \sum_{a,i} m_a^i\kappa_{ij}m_a^j 
    = -\frac{3}{\kappa} \sum_{a,i} d_{ijk} m_a^i m_a^j t^k 
    = \frac{6}{\kappa} \sum_i t^i c_{2i}(W) 
    \nonumber\\
    &\leq \frac{6}{\kappa} \|{\bm t}\| \|c_2(T{\cal M})\|,
\label{eq:kinequality}
\end{align}
where we use the supersymmetry condition (\ref{eq:Dterm}) in the first step and the 
tadpole cancellation condition (\ref{eq:tad}) in the last step. Here, $c_{2i}(W) = -\frac{1}{2} \sum_{a,j,k} d_{ijk}m_a^j m_a^k$ and $\|\cdot \|$ denotes the Euclidean norm. 
It suggests the following redefinition of the moduli metric:
\begin{align}
    \widetilde{G}_{ij} = \frac{\kappa}{6\|{\bm t}\|} G_{ij},
\end{align}
from which the inequality is rewritten as 
\begin{align}
    0 &< \sum_a m_a^i \widetilde{G}_{ij} m_a^j \leq \|c_2(T{\cal M})\|.
\end{align}
Furthermore, when we define ${\rm Eigen}(\widetilde{G}_{ij})\biggl|_{\rm min} = \lambda_{\rm min}$, 
the flux vector ${\bm m_a}$ is bounded from above by
\begin{align}
    \sum_a |{\bm m}_a|^2 \leq \frac{\|c_2(T{\cal M})\|}{\lambda_{\rm min}},
\end{align}
as derived in Ref. \cite{Constantin:2015bea}. 
For instance, in the case of single K\"ahler modulus, i.e., $h^{1,1}=1$, the modulus metric is given by
\begin{align}
    G_{11} 
    &=-3\left( \frac{\kappa_{11}}{\kappa} - \frac{3\kappa_1\kappa_1}{2\kappa^2}\right)
    = \frac{3}{2(t^1)^2}
\end{align}
with
\begin{align}
    \kappa = d_{111}(t^1)^3,\qquad \kappa_1 = d_{111}(t^1)^2,\qquad
    \kappa_{11}=d_{111}t^1.
\end{align}
Thus, the flux quanta are bounded from above as
\begin{align}
    \sum_a (m_a^1)^2 \leq \frac{4c_{2}(T{\cal M})}{d_{111}}.
\end{align}

\medskip

Let us derive the maximal value of the flux quanta $m_a^i$ in a different approach of Ref. 
\cite{Constantin:2015bea}. 
By using the so-called K-theory condition (\ref{eq:K-theory}) which is rewritten in terms of $m_a^i$\footnote{Note that we focus on $c_1(W)=0$ as mentioned below Eq. (\ref{eq:K-theory}).}:
\begin{align}
    \sum_a m_a^i = 0
\end{align}
for all $i$, it turns out that
\begin{align}
\sum_a |{\bm m}_a|^2
&= \sum_{i=1}^{h^{1,1}}\sum_{b=2}^n |m_b^i|^2 + |{\bm m}_1|^2
\nonumber\\
&\geq \sum_{i=1}^{h^{1,1}} \frac{\left( m_2^i + \cdots + m_n^{i}\right)^2}{n-1}
+|{\bm m}_1|^2
\nonumber
= \sum_{i=1}^{h^{1,1}} \frac{\left( -m_1^i\right)^2}{n-1}
+|{\bm m}_1|^2
= \frac{n}{n-1}|{\bm m}_1|^2,
\end{align}
where we use the Cauchy-Schwarz inequality. 
Note that a similar inequality holds on the other $a$, that is,
\begin{align}
    |{\bm m}_a|^2 \leq \frac{n-1}{n} \frac{\|c_2(T{\cal M})\|}{\lambda_{\rm min}},
    \label{eq:ma_inequality}
\end{align}
for all $a$. 
Thus, when we define $m_{\rm max}$ satisfying $|m_a^i| \leq {\rm max}_{a,i}(m_a^i)=:|m_{\rm max}|$ for all $a,i$, 
we arrive at 
\begin{align}
    |m_{\rm max}|^2 \leq \frac{n-1}{n} \frac{\|c_2(T{\cal M})\|}{\lambda_{\rm min}} - (h^{1,1}-1), 
\end{align}
where we assume $m_a^i\neq 0$ for all $i$, but it is possible to derive $m_a^i =0$ for some $i$. 
In this case, the term $h^{1,1}-1$ in the above equation will be modified. 
When we evaluate the index of chiral modes, we adopt the above conservative bound of the flux quanta 
throughout this paper. 

\medskip

We are now ready to derive the upper bound on the Atiyah-Singer index of chiral zero modes. 
Since the index of $SU(5)$ matter multiplets is determined by Eq. (\ref{eq:Index_SU(5)}), 
we find the following inequality:
\begin{align}
        |N_{\rm gen}| &= \biggl|\sum_{a,i,j,k} \frac{d_{ijk}}{6}m_a^i m_a^j m_a^k \biggl|
        \leq \frac{|m_{\rm max}|}{3} \|c_{2}(W)\|
        \leq \frac{|m_{\rm max}|}{3} \|c_{2}(T{\cal M})\|,
\end{align}
where we use the tadpole cancellation in the last step. 
Thus, the index is bounded from above by the maximal value of the flux quanta $m_{\rm max}$ 
and the second Chern number of the tangent bundle of CY. 
The authors of Ref. \cite{Constantin:2015bea} derived the index bounded by the CY volume itself, 
but we check that our finding bound is stronger than the known bound for all favorable complete intersection CY threefolds (CICYs). 

\medskip

Here and in what follows, we focus on CICYs defined in the ambient space $\mathbb{P}^{n_1}\times \cdots \times \mathbb{P}^{n_m}$. The CICYs are specified by the $m\times R$ configuration matrix \cite{Candelas:1987kf,Candelas:1987du}:
\begin{align}
\begin{matrix}
\mathbb{P}^{n_1}\\
\mathbb{P}^{n_2}\\
\vdots\\
\mathbb{P}^{n_m}\\
\end{matrix}
\begin{bmatrix}
q_1^1 & q_2^1 & \cdots & q_R^1\\
q_1^2 & q_2^2 & \cdots & q_R^2\\
\vdots & \vdots & \ddots & \vdots\\
q_1^m & q_2^m & \cdots & q_R^m\\
\end{bmatrix}
\end{align} 
where the positive integers $q^{\cal I}_r$ $(r=1,\cdots, R)$ denote the multi-degree of $R$ homogeneous polynomials on the ambient space $\mathbb{P}^{n_1}\times \cdots \times \mathbb{P}^{n_m}$ 
with the homogeneous coordinates of $\mathbb{P}^{n_{\cal I}}$ being $x^{\cal I}_\alpha$ for ${\cal I}=1,\cdots, m$ and $\alpha=0,\cdots,n_{\cal I}$. 
Since the first Chern class of the tangent bundle can be zero under $\sum_{r=1}^R q_r^{\cal I} =n_{\cal I}+1$ for all ${\cal I}$, the CICYs are defined on the common zero locus of these $R$ polynomials.\footnote{The dimension of CICYs is determined by $\sum_{{\cal I}=1}^m n_{\cal I}-R$.} 
In particular, we focus on all "favorable" CICYs among complete intersection CYs, 
where the "favorable" means that the second cohomology of CY descends from that of ambient space.\footnote{
See for the CICY list, Ref. \cite{Lukas}.} 

\medskip

\begin{figure}[H]
\begin{minipage}{0.5\hsize}
  \begin{center}
  \includegraphics[height=80mm]{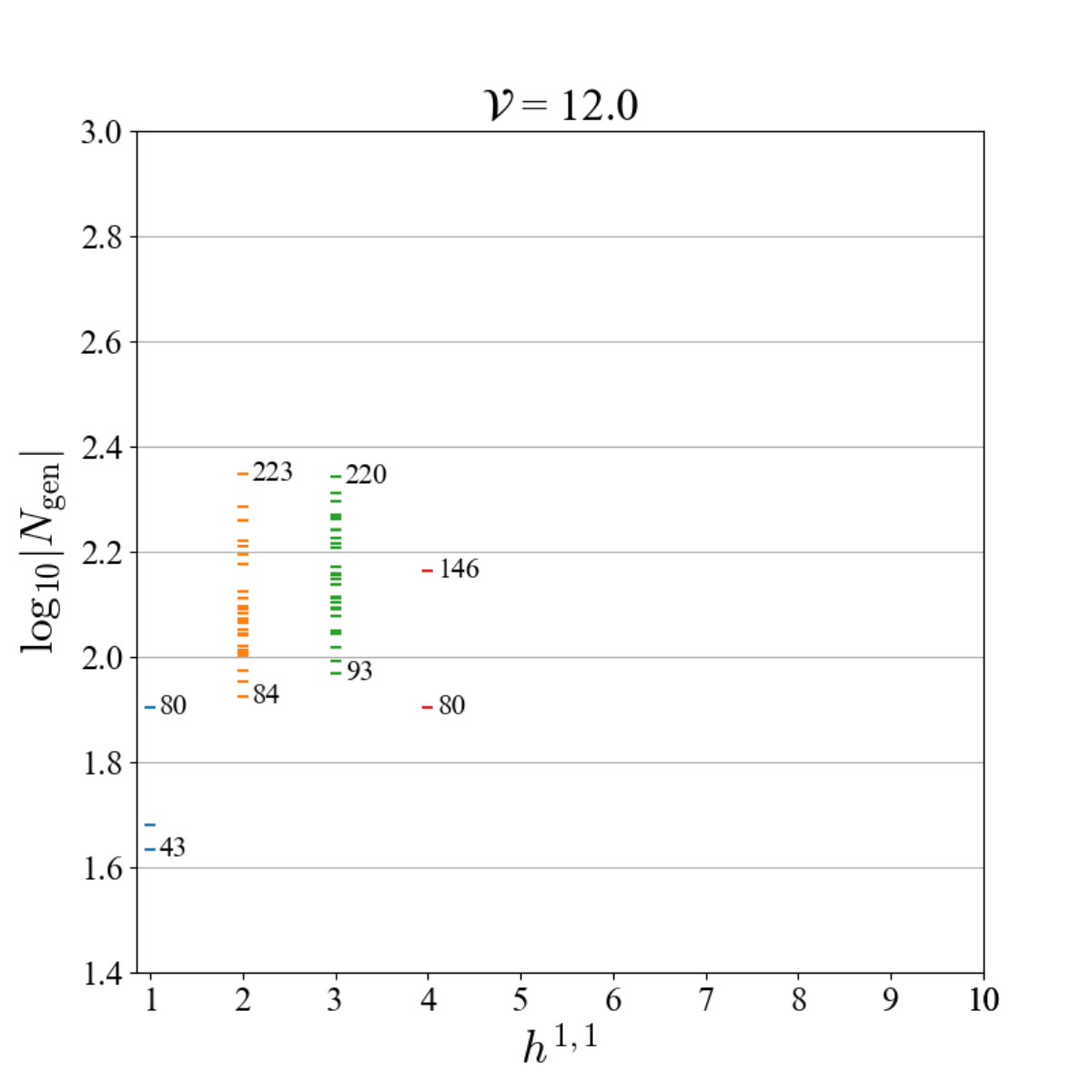}
  \end{center}
 \end{minipage}
 \begin{minipage}{0.5\hsize}
  \begin{center}
   \includegraphics[height=80mm]{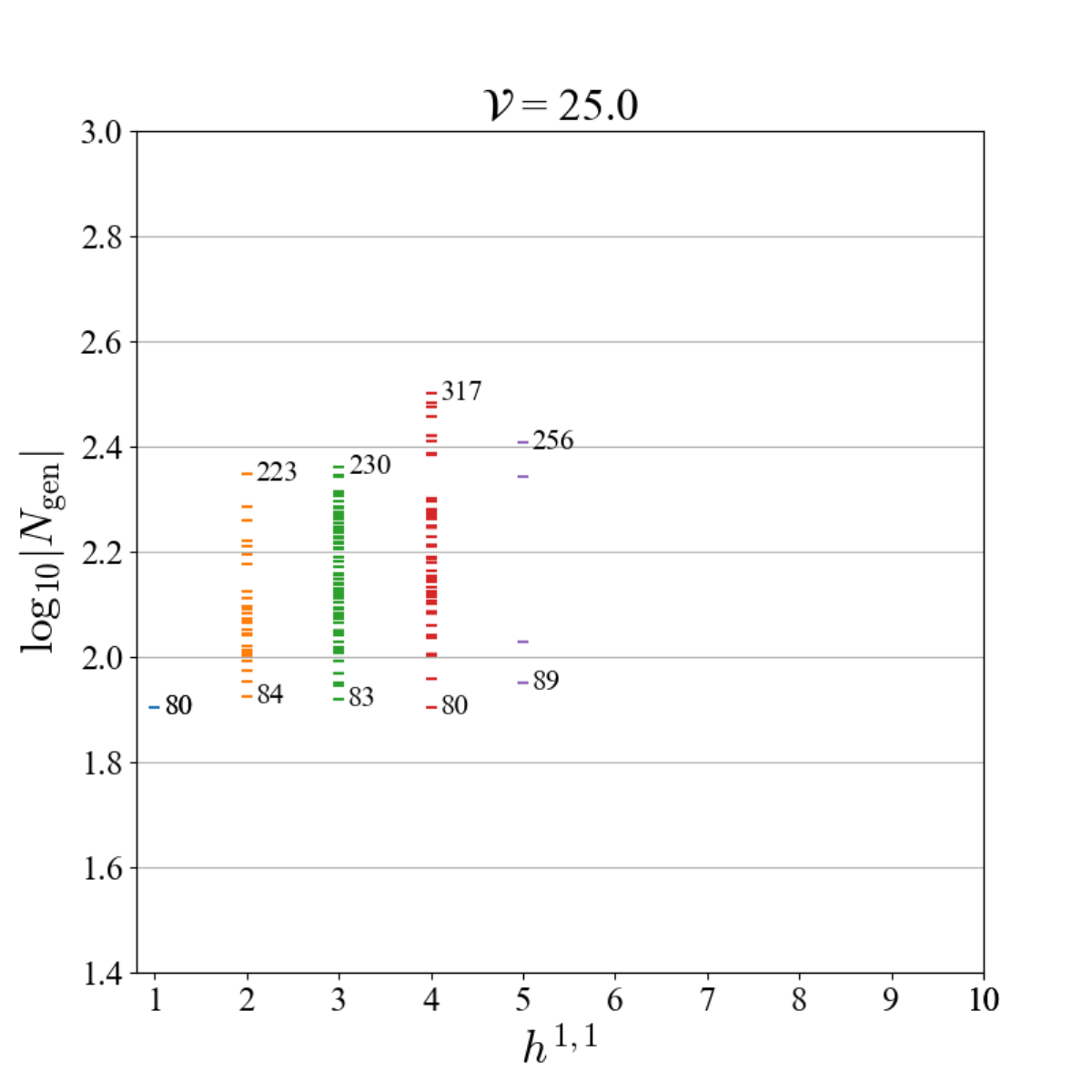}
  \end{center}
 \end{minipage}
  \caption{The upper bound on the chiral index in $SU(5)$ GUT on upstairs CICYs ${\cal M}$.}
\label{fig:SU(5)}
\end{figure}

To calculate the upper bound on the index, we have to calculate $\lambda_{\rm min}$ 
which depends on the values of moduli $t^i$. 
In our analysis, we adopt the universal value of moduli fields, $t:=t^i$ for all $i$ whose value is fixed by a given CY volume ${\cal V}=12$ or ${\cal V}=25$. 
By using the explicit values of CY topological data, we plot the index in Fig. \ref{fig:SU(5)}, where the CY volume is fixed as ${\cal V}=12$ in the left panel and ${\cal V}=25$ in the right panel, respectively. 
We find that there are no viable models on the large $h^{1,1}$ due to the fact that 
the volume of favorable CICYs is bounded from above by the phenomenological requirement (\ref{eq:CYvolume_bound}).

\begin{figure}[H]
\begin{minipage}{0.5\hsize}
  \begin{center}
  \includegraphics[height=80mm]{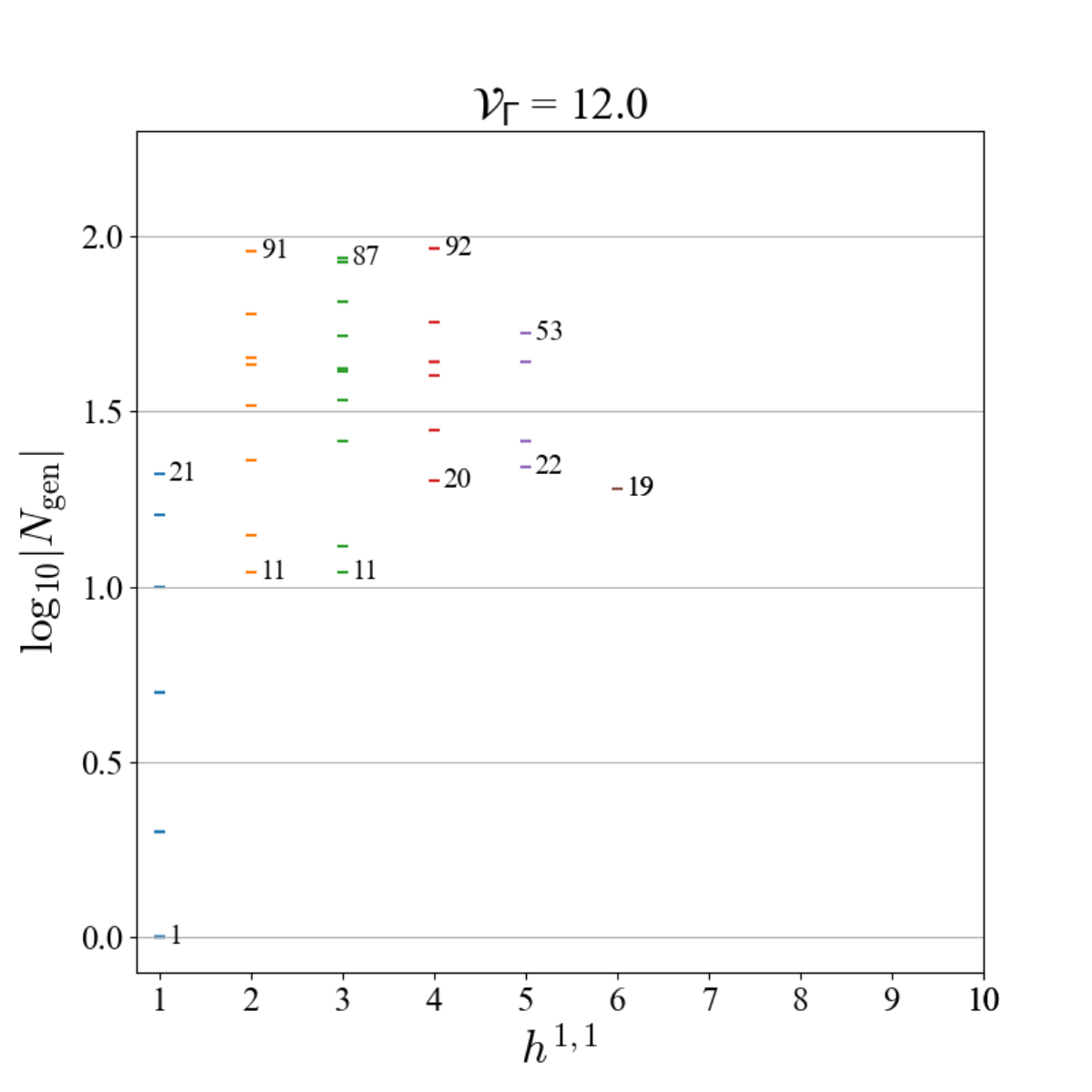}
  \end{center}
 \end{minipage}
 \begin{minipage}{0.5\hsize}
  \begin{center}
   \includegraphics[height=80mm]{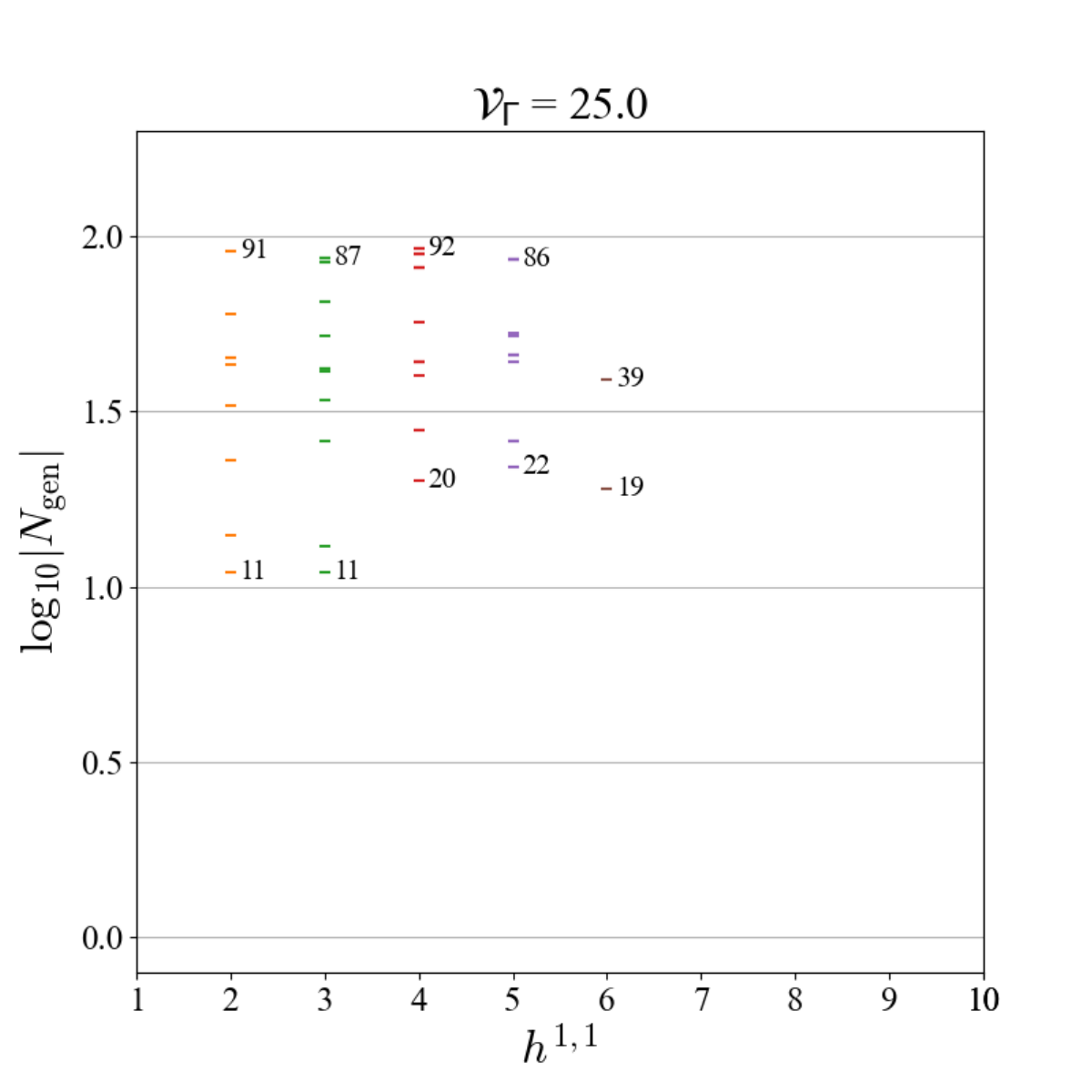}
  \end{center}
 \end{minipage}
  \caption{The upper bound on the chiral index in $SU(5)$ GUT on quotient CICYs ${\cal \widetilde{M}}$.}
\label{fig:SU(5)_quotient}
\end{figure}

So far, we have focused on the so-called "upstairs" CICYs. 
However, it was known that some of the CICYs admit a freely-acting discrete symmetry group $\Gamma$; one can define the quotient CICYs ${\cal \widetilde{M}}={\cal M}/\Gamma$~\cite{Candelas:1987kf,Candelas:1987du} on which one can turn on Wilson lines (see, Ref. \cite{Braun:2010vc} for the classification of $\Gamma$).  
Remarkably, such discrete quotients decrease the index:
\begin{align}
    N_{\rm gen}({\cal \widetilde{M}}) = \frac{N_{\rm gen}({\cal M})}{|\Gamma|},
\end{align}
although the topological quantities of quotient CICYs are different from before. 
Here, we consider a $\Gamma$-equivariant structure for the vector bundle $V$. 
For instance, the CY volume also reduces to ${\cal V}_{\Gamma} := {\cal V}/|\Gamma|$. For more details, see, e.g., Ref. \cite{Anderson:2009mh}. 
In Fig. \ref{fig:SU(5)_quotient}, we plot the chiral index on 195 quotient CICYs satisfying the volume constraint (\ref{eq:Volume_bound}). 
Similarly, the moduli values are chosen as $t:=t^i$ for all $i$ with 
the CY volume ${\cal V}_{\Gamma}=12$ in the left panel and ${\cal V}_{\Gamma}=25$ in the right panel of Fig. \ref{fig:SU(5)_quotient}, respectively. 
It turns out that the chiral index suppressed by $|\Gamma|$ satisfies $N_{\rm gen}\leq 100$ on quotient CY threefolds. 

It is interesting to apply our methods for finding a bound on the index to other classes of CY models. 
We study the CY threefolds in the Kreuzer-Skarke database \cite{Kreuzer:2000xy}, whose 
topological data was statistically analyzed in Ref. \cite{Demirtas:2018akl}. 
Note that we focus on favorable CY threefolds in the Kreuzer-Skarke database as in Ref. \cite{Demirtas:2018akl}. As seen in figure 1 of Ref. \cite{Demirtas:2018akl} with the large $h^{1,1}$ region, 
the nonvanishing triple intersection number behaves as
\begin{align}
    \#(d_{ijk}\neq 0) \gtrsim 6.5 h^{1,1} + 25,
\label{eq:dijk_toricCY}
\end{align}
for $h^{1,1} \gtrsim 25$\footnote{In contrast to Ref. \cite{Demirtas:2018akl}, 
we focus on the CY threefolds rather than their mirror duals.}. 
By introducing a normalized vector $\mathsf{t}^i$, we rewrite the K\"ahler moduli 
\begin{align}
    t^i = \mathsf{t}^i \|{\bm t}\|,
\end{align}
with $\|\mathsf{t} \|=1$, which scales as $\mathsf{t}^i \simeq (h^{1,1})^{-1/2}$ \cite{Demirtas:2018akl,Plauschinn:2021hkp}. 
Together with Eq. (\ref{eq:dijk_toricCY}), the CY volume in the geometrical regime scales as
\begin{align}
    {\cal V}
    \gtrsim  (h^{1,1})^{-1/2} \|{\bm t}\|^3 > (h^{1,1})^{-1/2},
\end{align}
which will be consistent with the bound on CY volume (\ref{eq:CYvolume_bound}) required in the GUT.

\medskip

Furthermore, one can derive the non-trivial bound on the CY topological quantities from the 
tadpole cancellation. 
By utilizing the scaling of $\kappa_i \simeq (h^{1,1})^{-1} \| {\bm t} \|^2$ and the mostly diagonal $\kappa_{ij}$ \cite{Demirtas:2018akl,Plauschinn:2021hkp}, i.e., $\kappa_{ii}\simeq (h^{1,1})^{-1/2} \|{\bm t} \|$, the sum of the flux vector is bounded from below by
\begin{align}
    \frac{\kappa}{3} \sum_{a,i,j} m_a^i G_{ij} m_a^j \geq (h^{1,1})^{1/2} \|{\bm t}\|,
\end{align}
with $h^{1,1}\gg 1$. 
Combining with Eq. (\ref{eq:kinequality}), we arrive at
\begin{align}
    \|c_2(T{\cal M})\| \geq \frac{(h^{1,1})^{1/2}}{2}.
\end{align}
Thus, the second Chern number of CY will be bounded by the hodge number due to the 
tadpole cancellation. 
Note that we have focused on favorable CYs, but it will be interesting to figure out the behavior of the chiral index on non-favorable CYs, which is left for future work.

\section{SO(32) heterotic string}
\label{sec:SO(32)}

In this section, we discuss the bound on the chiral index in the framework of 
$SO(32)$ heterotic string theory on smooth CY threefolds with line bundles.\footnote{For the model building on the toroidal background, see, e.g., Ref. \cite{Abe:2015mua}.}  
In particular, we focus on the Pati-Salam group in Sec. \ref{sec:pati-salam} 
and a direct flux breaking, so-called hypercharge flux breaking in Sec. \ref{sec:HC}. 

\subsection{Pati-Salam}
\label{sec:pati-salam}

Following Ref. \cite{Otsuka:2018oyf}\footnote{For MSSM-like models in $SO(32)$ heterotic string theory, see, Ref. \cite{GrootNibbelink:2015lme}.}, the Pati-Salam gauge group can be embedded into $SO(32)$ 
as follows:
\begin{align}
    SO(32) &\rightarrow SO(16) \times SO(16)_{\rm hid}  
    \nonumber\\
    &\rightarrow 
    SU(4)_C \times SU(2)_L \times SU(2)_R \times \Pi_{c=1}^2 U(1)_c  \times U(1)'\times SO(16)_{\rm hid},
\end{align}
where the Cartan directions of $SO(32)$, $H_a$ with $a=1,2,...,16$, are chosen as 
$H_1-H_2$, $H_1+H_2 - 2H_3$ and $H_1+H_2 + H_3 -3H_4$ for $SU(4)_C$, $H_5-H_6$ for $SU(2)_L$, 
$H_4+H_5$ for $SU(2)_R$, and $H_4-H_5$ for $U(1)'\simeq S(U(1)_4\times U(1)_5)$, respectively. 
On the $H_a$ basis, we choose the $SO(32)$ roots $(\underline{\pm 1, \pm 1, 0,\cdots, 0})$, where 
the underline represents all the possible permutations. $U(1)$s are chosen as
\begin{align}
U(1)_1&: (0,0,0,0,1,1,0,0; 0,\cdots,0),
\nonumber\\
U(1)_2&: (1,1,1,1,0,0,0,0; 0,\cdots,0),
\nonumber\\
U(1)_3&: (1,1,1,-3,0,0,0,0; 0,\cdots,0),
\nonumber\\
U(1)_4&: (0,0,0,0,0,0,1,0;0,\cdots,0),
\nonumber\\
U(1)_5&: (0,0,0,0,0,0,0,1; 0,\cdots,0),
\end{align}
respectively. 
The adjoint representation of $SO(32)$ splits into the spectrum in the visible and hidden sectors, 
as shown in Table \ref{tab:PS}, where the hypercharge direction is specified as
\begin{align}
    U(1)_Y = \frac{1}{6} U(1)_3 + \frac{1}{2}U(1)_4 + \frac{1}{2}U(1)_5.
\end{align}
\begin{table}[htp]
\centering
\small
  \begin{tabular}{|c|c|c|c|} \hline
    $\left(SU(4)_C\times SU(2)_L\times SU(2)_R\right.$ & $\left(SU(3)_C\times SU(2)_L\times SO(16)\right.$ 
    & & \\
    $\left.\times SO(16)\right)_{U(1)_1,U(1)_2,U(1)_4, U(1)_5}$ & $\left.\right)_{U(1)_1,U(1)_2,U(1)_3,U(1)_4,U(1)_5}$  
    & Matter &Index\\ \hline \hline
    \multicolumn{4}{|c|}{Visible sector}\\    \hline
    $(4,2,1,1)_{1,1,0,0}$ & $(3,2,1)_{1,1,1,0,0}$ & $Q_1$ & $\chi({\cal M},  l_1\otimes l_2)$ 
  \\
     & $(1,2,1)_{1,1,-3,0,0}$ & $L_1$ &    
    \\ \hline
    $(4,2,1,1)_{-1,1,0,0}$ & $(3,2,1)_{-1,1,1,0,0}$ & $Q_2$ & $\chi({\cal M},  l_1^{-1}\otimes l_2)$  
    \\
    & $(1,2,1)_{-1,1,-3,0,0}$ & $L_2$ &   
    \\ \hline
    $(15,1,1,1)_{0,0,0,0}$ & $({\bar 3},1,1)_{0,0,-4,0,0}$ & $u_{R_1}^c$ & $\chi({\cal M}, {\cal O}_{\cal M})$  
    \\ \hline
    $(6,1,1,1)_{0,2,0,0}$ & $({\bar 3},1,1)_{0,2,2,0,0}$ & $d_{R_1}^c$ & $\chi({\cal M},  l_2^2)$  
    \\ 
     & $(3,1,1)_{0,2,-2,0,0}$ & $\bar{d}_{R_2}^c$ &   
    \\ \hline
     $(1,1,1,1)_{2,0,0,0}$ & $(1,1,1)_{2,0,0,0,0}$ &  $n_1$ & $\chi({\cal M},  l_1^{2})$  
        \\ \hline
    $(\bar{4}, 1,2,1)_{0,-1,-1,0}$ & $({\bar 3},1,1)_{0,-1,-1,-1,0}$ & $u_{R_2}^{c\,\,4}$ & $\chi({\cal M}, l_2^{-1} \otimes l_4^{-1})$   
    \\ 
       & $(1,1,1)_{0,-1,3,0,1}$ &  $e_{R_1}^{c\,\,5}$ &    
    \\ 
        & $(1,1,1)_{0,-1,3,-1,0}$ & $n_{2}^{c\,4}$ &    
    \\ 
    & $({\bar 3},1,1)_{0,-1,-1,0,1}$ & $d_{R_3}^{c\,\,5}$ &   
    \\ \hline
    $(\overline{4}, 1, 2, 1)_{0,-1,1,0}$  & $({\bar 3},1,1)_{0,-1,-1,1,0}$ & $d_{R_3}^{c\,\,4}$ & $\chi({\cal M}, l_2^{-1}\otimes l_4)$   
    \\ 
       &$(1,1,1)_{0,-1,3,1,0}$ &  $e_{R_1}^{c\,\,4}$ &    
    \\ 
        & $(1,1,1)_{0,-1,3,0,-1}$ & $n_{2}^{c\,5}$ &    
    \\ 
    & $({\bar 3},1,1)_{0,-1,-1,0,-1}$ & $u_{R_2}^{c\,\,5}$ &    
    \\ \hline
    $(1,2,2,1)_{1,0,-1,0}$  & $(1,2,1)_{1,0,0,-1,0}$ & $L_3^4$ & $\chi({\cal M},  l_1\otimes l_4^{-1})$  
    \\ 
      & $(1,2,1)_{1,0,0,0,1}$ & $\bar{L}_4^5$ &    
    \\ \hline 
    $(1,2,2,1)_{1,0,1,0}$  & $(1,2,1)_{1,0,0,0,-1}$ &  $L_3^5$ &  $\chi({\cal M},  l_1\otimes l_4)$  
    \\ 
      & $(1,2,1)_{1,0,0,1,0}$ & $\bar{L}_4^4$ &    
    \\ \hline    
    $(1,1,3,1)_{0,0,0,0}$ & $(1,1,1)_{0,0,0,1,1}$ & $e_{R_2}^{c\,\,45}$ & $\chi({\cal M}, {\cal O}_{\cal M})$   
    \\ \hline
     $(1,1,1,1)_{0,0,2,0}$ & $(1,1,1)_{0,0,0,1,-1}$ & $n_{3}^{c\,45}$ &  $\chi({\cal M}, l_4^{2})$   
    \\ \hline
    \multicolumn{4}{|c|}{Hidden sector}\\    \hline
    $(\bar{4},1,1,16)_{0,-1,0,0}$ & $({\bar 3},1,16)_{0,-1,-1,0,0}$ & $-$ & $\chi({\cal M}, l_2^{-1})$   
    \\
      & $(1,1,16)_{0,-1,3,0}$ & $-$ &    
    \\ \hline    
    $(1,2,1,16)_{1,0,0,0}$ & $(1,2,16)_{1,0,0,0,0}$ & $-$ & $\chi({\cal M}, l_1)$   
    \\  \hline
    $(1,1,2,16)_{0,0,1,0}$ & $(1,1,16)_{0,0,0,1,0}$ & $-$ & $\chi({\cal M}, l_4)$   
    \\ 
     & $(1,1,16)_{0,0,0,0,-1}$ & $-$ &    
    \\ \hline    
     $(1,1,1,120)_{0,0,0,0}$ & $(1,1,120)_{0,0,0,0,0}$ &  $-$ & $\chi({\cal M}, {\cal O}_{\cal M})$   
    \\ \hline
  \end{tabular}
  \caption{Massless spectrum and corresponding index under $\left(SU(4)_C\times SU(2)_L\times SU(2)_R\times SO(16)\right)_{U(1)_1,U(1)_2,U(1)_4,U(1)_5}$ in the first column and $\left(SU(3)_C\times SU(2)_L\times SO(16)\right)_{U(1)_1,U(1)_2,U(1)_3,U(1)_4,U(1)_5}$ in the second column, respectively. 
  The $U(1)$ charges are specified by the subscript indices. Here, we represent $U(1)_{4,5}$ charges rather 
  than $U(1)'$, although both are correlated with $l_5\simeq l_4^{-1}$.}
  \label{tab:PS}
\end{table}
\normalsize

To avoid the existence of chiral exotic modes, we require
\begin{align}
    \chi (l_1) = \chi(l_2) = \chi(l_4) = -\chi(l_5)=0.
    \label{eq:PS_condition1}
\end{align}
The tadpole cancellation condition is given by
\begin{align}
    c_2(T{\cal M}) &\geq -{\rm ch}_2(W) = -2c_1^2(l_1) - 4c_1^2(l_2) - c_1^2(l_4)  - c_1^2(l_5),
    \label{eq:PS_tadpole}
\end{align}
with
\begin{align}
    c_1^2(l_a) = \sum_{j,k}d_{ijk}m_a^j m_a^k.
\end{align}

By identifying left-handed quarks $\{Q_1, Q_2\}$, right-handed quarks $\{u_{R_2}^{c\,\,4}, u_{R_2}^{c\,\,5}\}$ and vector-like quarks under the SM gauge group $d_{R_1}^c$, 
their generation numbers are counted by
\begin{align}
    N_{\rm gen}(Q)&:= N_{\rm gen}(Q_1) + N_{\rm gen}(Q_2)=\chi (l_1 \otimes l_2) + \chi (l_1^{-1} \otimes l_2), 
    \nonumber\\
    N_{\rm gen}(u_{R}^{c})&:= N_{\rm gen}(u_{R_2}^{c\,\,4}) + N_{\rm gen}(u_{R_2}^{c\,\,5})=\chi (l_2^{-1} \otimes l_4^{-1}) + \chi (l_2^{-1} \otimes l_4),
    \nonumber\\
    N_{\rm gen}^{\rm (vec)} &:= N_{\rm gen}(d_{R_1}^c)= \chi(l_2^{2}),
\end{align}
respectively. 
Note that the index of each field is calculated by using the Atiyah-Singer index theorem:
\begin{align}
\chi(l_a^p) &= \frac{p^3}{6}\sum_{i,j,k}d_{ijk}m_a^i m_a^j m_a^k + \frac{p}{12}\sum_i m_a^i c_2^{i}(T{\cal M}) = \frac{p^3}{6}X_{aaa} + \frac{p}{12}\sum_i m_a^i c_2^{i}(T{\cal M}),
\nonumber\\
\chi (l_a^p \otimes l_b^q) &= \chi(l_a^p) +\chi(l_b^q) + \frac{1}{2}(pq^2X_{abb} +p^2q X_{aab}),
\end{align}
with
\begin{align}
    X_{aaa} = \sum_{i,j,k} d_{ijk}m_a^i m_a^j m_a^k,\qquad
    X_{abb} = \sum_{i,j,k} d_{ijk}m_a^i m_b^j m_b^k,\qquad
    X_{aab} = \sum_{i,j,k} d_{ijk}m_a^i m_a^j m_b^k.
\end{align}
Thus, the generation numbers are rewritten as
\begin{align}
    N_{\rm gen}(Q)&= 2\chi(l_2) +\frac{1}{2}\left( X_{122} + X_{112} - X_{122} + X_{112}\right)
    =  2\chi(l_2) + X_{112}
    =  X_{112},
    \nonumber\\
    N_{\rm gen}(u_{R}^{c})&= 2\chi(l_2^{-1}) +\frac{1}{2}\left( -X_{244} - X_{224} - X_{244} + X_{224}\right)
    =  -2\chi(l_2) - X_{244}
    =  - X_{244},
    \nonumber\\
    N_{\rm gen}^{({\rm vec})}&= \chi(l_2^{2}) = X_{222},
\end{align}
where we used Eq. (\ref{eq:PS_condition1}). 
The index of left- and right-handed quarks should be equal, that is, $N_{\rm gen}^{\rm (quark)}= N_{\rm gen}(Q)=N_{\rm gen}(u_{R}^{c})$. 
In the following, we will derive the upper bound on these Atiyah-Singer indices. 

\medskip

From the equality:
\begin{align}
    2N_{\rm gen}(Q) + 4N_{\rm gen}^{\rm (vec)} + 2N_{\rm gen}(u_{R}^{c})
    &= 2X_{112} + 4X_{222} -2 X_{244} 
    \nonumber\\
    &=m_2^k \left(2d_{ijk}m_1^i m_1^j + 4 d_{ijk}m_2^i m_2^j +2d_{ijk}m_4^i m_4^j - 4d_{ijk}m_4^i m_4^j\right) 
    \nonumber\\
    &= -\sum_k m_2^k c_2(W)_k + 4N_{\rm gen}(u_{R}^{c}), 
\end{align}
we arrive at the following formula:
\begin{align}
    |N_{\rm gen}^{\rm (vec)}|
    = \frac{|\sum_k m_2^k c_2(W)_k|}{4} \leq \frac{|m_{\rm max}| \|c_2(W)\|}{4} \leq \frac{|m_{\rm max}| \|c_2(T{\cal M})\|}{4},
\end{align}
where we use the tadpole cancellation condition (\ref{eq:PS_tadpole}) and $|m_2^k|\leq |m_{\rm max}|$ 
for all $k$. 
Thus, the generation number of vector-like quarks is constrained by 
\begin{align}
     |N_{\rm gen}^{\rm (vec)}| \leq \frac{|m_{\rm max}| \|c_2(T{\cal M})\|}{4}.
\end{align}

\medskip

To identify Higgs fields, let us take $N_{\rm gen}(u_{R_2}^{c\,\,4})=0$ for simplicity.\footnote{We obtain the same bound in the case with $N_{\rm gen}(u_{R_2}^{c\,\,5})=0$.}
Since the following Yukawa couplings of quarks and leptons are allowed under the gauge symmetries of the low-energy effective action:
\begin{align}
&(Q_1, \bar{L}_3^5, u_{R_2}^{c\,\,5}),\qquad (Q_2, \bar{L}_4^5, u_{R_2}^{c\,\,5}),\qquad
\end{align}
we identify the up-type Higgs doublets as $\bar{L}_{3, 4}^5$, which are 
vector-like particles under the SM gauge group but are chiral under the extra $U(1)$ symmetries. 
The index of Higgs fields is then given by
\begin{align}
    N_{\rm gen}(H) &:= N_{\rm gen}(\bar{L}_4^5) + N_{\rm gen}(\bar{L}_3^5)= \chi(l_1 \otimes l_4^{-1}) + \chi(l_1^{-1} \otimes l_4^{-1})=-X_{114},
\end{align}
which is correlated with the quark index as
\begin{align}
    N_{\rm gen}(H) -2N_{\rm gen}(u_{R_2}^{c\,\,5}) + \frac{\sum_i m_4^i (- c_2(W)_i + c_2(T{\cal M})_i)}{2}=0.
\end{align}
Similarly, the tadpole cancellation condition provides the bound on the chiral index:
\begin{align}
    |N_{\rm gen}(H) - 2N_{\rm gen}^{\rm (quark)}| \leq \frac{|m_{\rm max}|}{2}\sum_s N_s.
\end{align}
If $N_{\rm gen}^{\rm (quark)}=-3$, the generation number of Higgs fields is bounded by the brane charge:
\begin{align}
    |N_{\rm gen}(H) +6| \leq \frac{|m_{\rm max}|}{2} \sum_s N_s.
\end{align}
Thus, an arbitrary number of Higgs fields is not allowed in the context of $SO(32)$ heterotic line bundle models.

\subsection{Hypercharge flux}
\label{sec:HC}

So far, we have focused on the visible sector, equipping the $SU(5)$ or Pati-Salam 
gauge symmetries. 
In this section, we deal with a direct flux breaking, so-called hypercharge flux 
breaking in heterotic string theory as discussed in Refs. \cite{Blumenhagen:2005ga,Blumenhagen:2005pm,Blumenhagen:2006ux,Blumenhagen:2006wj}. 
Since this approach does not require the existence of Wilson lines, 
it is applicable to simply-connected manifolds. 
However, it was pointed out that the $E_8\times E_8$ heterotic string with hypercharge flux 
does not lead to the SM spectra as a consequence of the large number of index conditions \cite{Anderson:2014hia}. 
This is the reason why we focus on the direct flux breaking scenario in the context of 
$SO(32)$ heterotic string theory.

\medskip

As systematically analyzed in Ref. \cite{Otsuka:2018rki}, 
the SM gauge group can be directly realized from $SO(16)$ gauge group:
\begin{align}
    SO(32) &\rightarrow SO(16) \times SO(16)_{\rm hid}  
    \nonumber\\
    &\rightarrow 
    SU(3)_C \times SU(2)_L \times \Pi_{a=1}^5 U(1)_a \times SO(16)_{\rm hid},
\end{align}
where the Cartan directions of $SO(32)$, $H_a$ with $a=1,2,...,16$, are chosen as 
$H_1-H_2$ and $H_2 - H_3$ for $SU(3)_C$, $H_4-H_5$ for $SU(2)_L$, respectively. 
The other $U(1)$s are chosen as
\begin{align}
U(1)_1&: (1,1,1,0,0,0,0,0;0,\cdots,0),
\nonumber\\
U(1)_2&: (0,0,0,1,1,0,0,0;0,\cdots,0),
\nonumber\\
U(1)_3&: (0,0,0,0,0,1,0,0;0,\cdots,0),
\nonumber\\
U(1)_4&: (0,0,0,0,0,0,1,0;0,\cdots,0),
\nonumber\\
U(1)_5&: (0,0,0,0,0,0,0,1;0,\cdots,0),
\end{align}
respectively. 
The adjoint representation of $SO(32)$ splits into the spectrum in the visible and hidden sectors, 
as shown in Table \ref{tab:hypercharge}, where the hypercharge direction is specified as
\begin{align}
U(1)_Y=-\frac{1}{6}U(1)_1 -\frac{1}{2}\biggl(U(1)_3 -U(1)_4 +U(1)_5\biggl).
\label{eq:U(1)Y_hyper}
\end{align}
To avoid the existence of chiral exotic modes, we require
\begin{align}
\chi (l_1)=\chi (l_2)=\chi (l_3)=\chi (l_4)=\chi (l_5)=\chi (l_1^{-2})=0.
    \label{eq:HC_condition1}
\end{align}
The tadpole cancellation condition is given by
\begin{align}
    c_2(T{\cal M}) &\geq -{\rm ch}_2(W) = -3c_1^2(l_1) - 2c_1^2(l_2) - c_1^2(l_3) - c_1^2(l_4)  - c_1^2(l_5).
    \label{eq:HC_tadpole}
\end{align}
The generation numbers of left-handed and right-handed quarks/leptons are counted by
\begin{align}
N_{\rm gen}(Q)&:= \chi (l_1^{-1}\otimes l_2) + \chi (l_1^{-1}\otimes l_2^{-1}), \nonumber\\
N_{\rm gen}(L)&:= \sum_{s=\pm 1}\chi (l_2^{s}\otimes l_3^{-1})+ \sum_{s=\pm 1}\chi (l_2^{s}\otimes l_4^{1})
+\sum_{s=\pm 1}\chi (l_2^{s}\otimes l_5^{1}), \nonumber\\
N_{\rm gen}(u^c)&:= \chi (l_1\otimes l_3) + \chi (l_1\otimes l_4^{-1}) + \chi (l_1\otimes l_5), \nonumber\\
N_{\rm gen}(d^c)&:= \chi (l_1\otimes l_3^{-1}) + \chi (l_1\otimes l_4) + \chi (l_1\otimes l_5^{-1}), \nonumber\\
N_{\rm gen}(e^c)&:= \chi (l_3^{-1}\otimes l_4)+ \chi (l_3^{-1}\otimes l_5^{-1}) + \chi (l_4\otimes l_5^{-1}),
\label{eq:HC}
\end{align}
respectively. 
\begin{table}[H]
\centering
\small
  \begin{tabular}{|c|c|c|} \hline
    Matter & Repr. & Index  \\ \hline \hline
    \multicolumn{3}{|c|}{Visible sector}\\    \hline
    $Q_1$ & $(3,2)_{-1,1,0,0,0}$ & $\chi({\cal M},  l_1^{-1}\otimes l_2)$ 
     \\
    $Q_2$ & $(3,2)_{-1,-1,0,0,0}$ & $\chi({\cal M}, l_1^{-1}\otimes l_2^{-1})$  \\ \hline
    $L_1$ & $(1,2)_{0,1,1,0,0}$ & $\chi({\cal M},  l_2\otimes l_3)$ \\
    $L_2$ & $(1,2)_{0,-1,1,0,0}$ & $\chi({\cal M},  l_2^{-1}\otimes l_3)$ \\ 
    $L_3$ & $(1,2)_{0,1,0,-1,0}$ & $\chi({\cal M},  l_2\otimes l_4^{-1})$ \\
    $L_4$ & $(1,2)_{0,-1,0,-1,0}$ & $\chi({\cal M},  l_2^{-1}\otimes l_4^{-1})$ \\ 
    $L_5$ & $(1,2)_{0,1,0,0,1}$ & $\chi({\cal M},  l_2\otimes l_5)$ \\
    $L_6$ & $(1,2)_{0,-1,0,0,1}$ & $\chi({\cal M},  l_2^{-1}\otimes l_5)$ \\ \hline
    $u_{R_1}^c$ & $({\bar 3},1)_{1,0,1,0,0}$ & $\chi({\cal M},  l_1\otimes l_3)$ \\
    $u_{R_2}^c$ & $({\bar 3},1)_{1,0,0,-1,0}$ & $\chi({\cal M},  l_1\otimes l_4^{-1})$ \\
    $u_{R_3}^c$ & $({\bar 3},1)_{1,0,0,0,1}$ & $\chi({\cal M},  l_1\otimes l_5)$ \\
    \hline
    $d_{R_1}^c$ & $({\bar 3},1)_{1,0,-1,0,0}$ & $\chi({\cal M},  l_1\otimes l_3^{-1})$ \\
    $d_{R_2}^c$ & $({\bar 3},1)_{1,0,0,1,0}$ & $\chi({\cal M},  l_1\otimes l_4)$ \\
    $d_{R_3}^c$ & $({\bar 3},1)_{1,0,0,0,-1}$ & $\chi({\cal M},  l_1\otimes l_5^{-1})$ \\
    \hline
    $e_{R_1}^{c}$ & $(1,1)_{0,0,-1,1,0}$ & $\chi({\cal M}, l_3^{-1} \otimes l_4)$  \\ 
    $e_{R_2}^{c}$ & $(1,1)_{0,0,-1,0,-1}$ & $\chi({\cal M}, l_3^{-1} \otimes l_5^{-1})$  \\ 
    $e_{R_3}^{c}$ & $(1,1)_{0,0,0,1,-1}$ & $\chi({\cal M}, l_4 \otimes l_5^{-1})$  \\ 
    \hline
    $n_1^c$ & $(1,1)_{0,-2,0,0,0}$ & $\chi({\cal M},  l_2^{2})$ \\    
    $n_2^c$ & $(1,1)_{0,0,-1,-1,0}$ & $\chi({\cal M},  l_3^{-1}\otimes l_4^{-1})$ \\    
    $n_3^c$ & $(1,1)_{0,0,-1,0,1}$ & $\chi({\cal M},  l_3^{-1}\otimes l_5)$ \\    
    $n_4^c$ & $(1,1)_{0,0,0,1,1}$ & $\chi({\cal M},  l_4\otimes l_5)$ \\    
    \hline
    $\phi$  & $({\bar 3},1)_{-2,0,0,0,0}$ & $\chi({\cal M},  l_1^{-2})$ \\
    \hline
    \multicolumn{3}{|c|}{Hidden sector}\\    \hline
    $-$ & $(3,1,16)_{1,0,0,0}$ & $\chi({\cal M}, l_1 )$   \\
    $-$ & $(1,2,16)_{0,1,0,0,0}$ & $\chi({\cal M}, l_2)$   \\
    $-$ & $(1,1,16)_{0,0,1,0,0}$ & $\chi({\cal M}, l_3)$   \\
    $-$ & $(1,1,16)_{0,0,0,1,0}$ & $\chi({\cal M}, l_4)$   \\
    $-$ & $(1,1,16)_{0,0,0,0,1}$ & $\chi({\cal M}, l_5)$   \\
    $-$ & $(1,1,120)_{0,0,0,0,0}$ & $\chi({\cal M}, {\cal O}_{\cal M})$   \\ \hline
  \end{tabular}
  \caption{Massless spectrum and corresponding index under $\left(SU(3)_C\times SU(2)_L\times SO(16)\right)_{U(1)_1,U(1)_2,U(1)_3,U(1)_4,U(1)_5}$. 
  Here, the subscript indices label the $U(1)_{1,2,3,4,5}$ charges.}
  \label{tab:hypercharge}
\end{table}

Note that the hypercharge gauge boson will couple to string axions through the St$\mathrm{\ddot{u}}$ckelberg couplings. 
Taking into account the St$\mathrm{\ddot{u}}$ckelberg couplings of K$\mathrm{\ddot{a}}$hler axions and dilaton axion, 
the hypercharge gauge boson $U(1)_Y = \sum_a f_a U(1)_a$ becomes massless under\footnote{For more details, see, Ref. \cite{Otsuka:2018rki}.} 
\begin{align}
&\sum_{a} {\rm tr}(T_a^2) f_a m_a^{(i)}=0, 
\label{eq:hyp1}
\\
&\sum_{a,b,c,d} {\rm tr}(T_aT_bT_cT_d)f_a X_{bcd}=0,
\label{eq:hyp2}
\end{align}
where $f_a$ is chosen as in Eq. (\ref{eq:U(1)Y_hyper}). 
Furthermore, flux quanta are constrained by the K-theory condition (\ref{eq:K-theory}):
\begin{align}
\sum_a {\rm tr}(T_a)m_a^i= 0,
\end{align}
with $n_a = {\rm tr}(T_a)$ in Eq. (\ref{eq:K-theory}). 
Together with the masslessness condition of the hypercharge gauge boson (\ref{eq:hyp1}), 
we obtain
\begin{align}
    m_1^i + m_2^i + m_4^i = 0,
\end{align}
which leads to the following equality\footnote{Here, we omit the summation over $i$.}:
\begin{align}
    m_2^ic_2(W)_i &= -m_2^i (3c_1^2(l_1)+2c_1^2(l_2) + c_1^2(l_3) + c_1^2(l_4) + c_1^2(l_5))_i 
    \nonumber\\
    &= -3X_{112} -2 X_{222} +(m_1^i + m_4^i)(c_1^2(l_3) + c_1^2(l_4) + c_1^2(l_5))_i 
    \nonumber\\
    &= - 3(N_{\rm gen}(Q_1)- N_{\rm gen}(Q_2)) -2 X_{222} +N_{\rm gen}(u^c) + N_{\rm gen}(d^c) 
    -m_4^i c_2(W)_i 
    \nonumber\\
    &\quad-m_4^i (3c_1^2(l_1) + 2c_1^2(l_2))_i
    \nonumber\\
    &= - 3(N_{\rm gen}(Q_1)- N_{\rm gen}(Q_2)) -2 X_{222} +N_{\rm gen}(u^c) + N_{\rm gen}(d^c) 
    -m_4^i c_2(W)_i
    \nonumber\\    
    &\quad +3(N_{\rm gen}(u_{R_2}^c) - N_{\rm gen}(d_{R_2}^c)) +2 (N_{\rm gen}(L_3) + N_{\rm gen}(L_4)).
\label{eq:equality1}
\end{align}
Here, we use Eq. (\ref{eq:HC_condition1}) and 
\begin{align}
     N_{\rm gen}(u_{R_2}^c) - N_{\rm gen}(d_{R_2}^c) &= - X_{114},
    \nonumber\\
    N_{\rm gen}(L_3) + N_{\rm gen}(L_4) &= -X_{224} ,
    \nonumber\\
    N_{\rm gen}(Q_1) - N_{\rm gen}(Q_2) &= X_{112},
    \nonumber\\
    N_{\rm gen}(u^c) + N_{\rm gen}(d^c) &= X_{133} + X_{144} + X_{155}.
\end{align}
Since $\sum_i m_1^i c_2(W)_i = - \sum_i (m_2^i + m_4^i)c_2(W)_i =0$ is satisfied when $N_{\rm gen}(Q) = N_{\rm gen}(u^c) = N_{\rm gen}(d^c)$, 
Eq. (\ref{eq:equality1}) is further simplified as
\begin{align}
&\sum_i m_2^i c_2(T{\cal M})_i
\nonumber\\
&=3(N_{\rm gen}(Q_1)- N_{\rm gen}(Q_2)) -2N_{\rm gen}(Q) 
-3(N_{\rm gen}(u_{R_2}^c) - N_{\rm gen}(d_{R_2}^c)) -2 (N_{\rm gen}(L_3) + N_{\rm gen}(L_4))
,
\end{align}
where we use $N_{\rm gen}(Q) = N_{\rm gen}(u^c) = N_{\rm gen}(d^c)$ and $2X_{222} = -\sum_i m_2^i c_2(T{\cal M})_i$ due to $\chi(l_2)=0$. 
Thus, if $N_{\rm gen}(Q) = N_{\rm gen}(Q_1)$ with $N_{\rm gen}(Q_2)=0$, we obtain a similar bound for the sum of the index of chiral particles, as discussed in previous sections:
\begin{align}
     |N_{\rm gen}(Q) -3(N_{\rm gen}(u_{R_2}^c) - N_{\rm gen}(d_{R_2}^c)) -2 (N_{\rm gen}(L_3) + N_{\rm gen}(L_4))| \leq |m_{\rm max}| \|c_2(T{\cal M})\|.
\end{align}
Note that the value of indices of $u_{R_2}^c, d_{R_2}^c$ and $L_{3,4}$ depends on a concrete model, but 
it is typically of ${\cal O}(N_{\rm gen}(Q))$.

\section{Type II string}
\label{sec:typeII}

In this section, we move to the Type II orientifold compactifications. 
In particular, we focus on Type IIB orientifold compactifications with 
O3- and O7-planes, but the analysis can be extended to the T-dual Type IIA intersecting D-brane models. 

\medskip

On $N_a$ stacks of magnetized D7-branes wrapping holomorphic divisors $\Gamma_{\rm D7}$ of Calabi-Yau threefolds ${\cal M}$ \footnote{From the calibration condition of D7-branes, four-cycles $\Gamma_{\rm D7}$ are considered the holomorphic divisor \cite{Marino:1999af}.}, world-volume fluxes induce the degenerate number of chiral zero modes. 
Similar to the previous sections, it is possible to embed $U(1)$ gauge fluxes into $U(N_a)$ on the D7-branes that do not lie on the O7-planes 
or $SO(2N_a)/Sp(2N_a)$ on the D7-branes that lie on the O7-planes. 
Specifically, we focus on $U(N_a)$ magnetized D7-branes. 
Following Ref. \cite{Blumenhagen:2008zz}, we choose the background gauge invariant field strength:
\begin{align}
    {\cal F}_a= (\bar{F}_a^0 + 2\pi \iota^\ast B_2 )T_0 + \sum_i\bar{F}_a^i T_i,
\end{align}
where $\iota^\ast$ denotes the pull-back from the CY threefolds ${\cal M}$ to $\Gamma_{\rm D7}$. 
Here, $T_0$ and $T_i$ denote the generators of diagonal $U(1)_a$ and the other traceless Abelian elements of $U(N_a)$, respectively. 
An orientifold projection $\Omega (-1)^{F_L}\sigma$ for $B_2$ and ${\cal F}$ is chosen as
\begin{align}
    \Omega (-1)^{F_L}B_2 = -B_2,\qquad
    \Omega (-1)^{F_L}{\cal F} = - {\cal F},
\end{align}
and the holomorphic involution $\sigma$ acts on the K\"ahler form and NS-NS two-form $B_2$ on ${\cal M}$:
\begin{align}
    \sigma^\ast J = + J,
    \qquad
    \sigma^\ast B_2^\pm = \pm B_2^\pm.
\end{align}
Thus, the following internal gauge invariant field strength will be relevant to the 
realization of chiral zero modes:
\begin{align}
    \overline{{\cal F}}^+ = -i (l_s^2 \bar{F} + 2\pi \iota^\ast B_2^+ \mathbb{I}).
\end{align}
Note that such background fluxes $\bar{F}$ are assumed to be holomorphic and preserve the supersymmetry:
\begin{align}
J \wedge \bar{F} =0    
\end{align}
on the divisor $\Gamma_{\rm D7}$, where $J$ denotes the K\"ahler form of ${\cal M}$. 
Furthermore, such magnetized D7-branes as well as O7-planes induce the D3-brane charge:
\begin{align}
    Q_{\rm D7}^a &= N_a \frac{\chi(D_a)}{24} + \frac{1}{8\pi^2}\int_{D_a} {\rm tr}{\cal F}_a^2,
    \nonumber\\
    Q_{\rm O7} &= \frac{\chi (D_{\rm O7})}{6} = \frac{1}{6}\int_{\cal M} c_2 (D_{\rm O7}) \wedge [D_{\rm O7}],
\end{align}
where $\chi(D_a)$ and $\chi(D_{\rm O7})$ denote the Euler characteristic of divisors $D_a$ and $D_{\rm O7}$, respectively. 
Then, the D3-brane tadpole cancellation conditions become
\begin{align}
    N_{\rm D3} + \frac{N_{\rm flux}}{2} + N_{\rm gauge} = \frac{\chi (Y_4)}{24},
\end{align}
where 
\begin{align}
    N_{\rm gauge} &= - \sum_a \frac{1}{8\pi^2}\int {\rm tr}{\cal F}_a^2,
    \nonumber\\
    \frac{\chi (Y_4)}{24} &= \frac{N_{O3}}{4} + \frac{\chi(D_{O7})}{12} + \sum_a N_a \frac{\chi(D_a)}{24}.
\end{align}
Here, we suppose an uplifting of Type IIB orientifolds to the F-theory compactifications on Calabi-Yau fourfolds $Y_4$ with Euler characteristics $\chi(Y_4)$. 

\medskip

As discussed in Secs. \ref{sec:E8} and \ref{sec:SO(32)}, we derive the upper bound on the magnetic flux quanta determining the number of chiral zero modes. 
Now that the gauge fluxes are rewritten by
\begin{align}
    N_{\rm gauge} &= -\frac{1}{2}\sum_a N_a \int_{D_a} c_1(l_a) \wedge c_1(l_a) 
    = -\frac{1}{2}\sum_a N_a \int_{\cal M} c_1(l_a) \wedge c_1(l_a) \wedge [D_a]
    \nonumber\\
    &= \frac{1}{2}\sum_a N_a m_a^2 \kappa_{aaa},
\end{align}
with $c_1(l_a) = m_a [D_a]$ and $\kappa_{aaa}:= -\int_{\cal M} [D_a]\wedge [D_a]\wedge [D_a]$, 
it is bounded from above by using the tadpole cancellation condition:
\begin{align}
\frac{{\rm min}_a (N_a\kappa_{aaa})}{2}\sum_a m_a^2 
\leq N_{\rm gauge} = \frac{\chi (Y_4)}{24} -  N_{\rm D3} - \frac{N_{\rm flux}}{2} \leq \frac{\chi (Y_4)}{24}.
\end{align}
Here, we use $N_{\rm flux}\geq 0$ due to the imaginary self-dual condition of three-form fluxes. 
The above expression is similar to Eq. (\ref{eq:ma_inequality}) discussed in the heterotic line bundle models. 
When we define $m_{\rm max}$ satisfying $|m_a| \leq {\rm max}_{a}(m_a):=|m_{\rm max}|$ for all $a$, 
we find that 
\begin{align}
    |m_{\rm max}|^2 \leq \frac{\chi (Y_4)}{12\,{\rm min}_a(N_a \kappa_{aaa})} - (h^{1,1}(D_a)-1), 
    \label{eq:mmax_bound_TypeII}
\end{align}
where we assume $m_a^i\neq 0$ for all $a$, but it is possible to derive $m_a =0$ for some $a$. 
Although the term $h^{1,1}(D_a)-1$ in the above equation will be modified in this case, 
we adopt the above conservative bound of the flux quanta in the following analysis.

\medskip

Let us discuss the chiral zero modes arising from intersections of two different stacks of magnetized D7-branes carrying holomorphic line bundles $l_a$ and $l_b$, respectively. 
In particular, we focus on matter fields in the bifundamental representation $(N_a, \bar{N}_b)$, as classified in the following two classes\footnote{Fore more details, see, e.g., Ref \cite{Blumenhagen:2008zz}.}: 
\begin{enumerate}
    \item Matter divisors

    If divisors of two D7-brane coincide with each other, i.e., $D:=D_a=D_b$, 
    the index of bifundamental matter fields is counted by the extension group \cite{Katz:2002gh}. 
    The chiral index is determined by
    \begin{align}
        I_{ab} 
        &= \int_{\cal M} [D] \wedge [D] \wedge (c_1(l_a) - c_1(l_b))
        \nonumber\\
        &= -(m_a - m_b) \kappa_{DDD},
    \end{align}
    with $c_1(l_{a,b}) = m_{a,b} [D]$, $\kappa_{DDD}:= -\int_{\cal M} [D]\wedge [D]\wedge [D]$. 
    Thus, the chiral index is bounded from above by
    \begin{align}
        |I_{ab}| \leq 2|m_{\rm max}| |\kappa_{DDD}| 
        \leq 2|\kappa_{DDD}|\left(\frac{\chi (Y_4)}{12\,{\rm min}_a(N_a \kappa_{aaa})} - (h^{1,1}_{-}(D_a^+)-1)\right),
    \end{align}
    where we use Eq. (\ref{eq:mmax_bound_TypeII}) in the last inequality. 

    \item Matter curves

    If divisors of two D7-brane are different but intersect over a curve $C$ of genus $g$, 
    the chiral index is determined by
    \begin{align}
        I_{ab}
        &= \int_{\cal M} [D_a] \wedge [D_b] \wedge (c_1(l_a) - c_1(l_b))
        \nonumber\\
        &= -m_a \kappa_{aab} + m_b \kappa_{abb},
    \end{align}
    with $c_1(l_{a}) = m_{a} [D_a]$, $c_1(l_{b}) = m_{b} [D_b]$, $\kappa_{aab}:= -\int_{\cal M} [D_a]\wedge [D_a]\wedge [D_b]$ and $\kappa_{abb}:= -\int_{\cal M} [D_a]\wedge [D_b]\wedge [D_b]$. 
    By using Eq. (\ref{eq:mmax_bound_TypeII}), we arrive at the upper bound on 
    the chiral index:
    \begin{align}
        |I_{ab}| &\leq 2|m_{\rm max}| |{\rm max}(\kappa_{aab}, \kappa_{abb})| 
        \nonumber\\
        &\leq 2|{\rm max}(\kappa_{aab}, \kappa_{abb})| \left(\frac{\chi (Y_4)}{12\,{\rm min}_a(N_a \kappa_{aaa})} - (h^{1,1}_{-}(D_a^+)-1)\right).
    \end{align}

\end{enumerate}

For illustrative purposes, let us focus on toroidal orientifolds. 
The distribution of flux quanta was explicitly discussed in Ref. \cite{Kai:2023ivp} in the context of Type IIB string theory on $T^6/(\mathbb{Z}_2\times \mathbb{Z}_2^\prime)$ orientifolds with $N_a$ stacks of magnetized D$(3+2n)$-branes and O-planes. 
When we turn on $U(1)_a$ magnetic fluxes on $(T^2)_i$:
\begin{align}
    \frac{m_a^i}{2\pi} \int_{(T^2)_i} F_a^i = n_a^i,
\end{align}
the wrapping numbers $m_a^i$ and magnetic fluxes $n_a^i$ lead to the net number of chiral zero modes between two stacks $a$ and $b$ of D-branes:
\begin{align}
    I_{ab}  = \Pi_{i=1}^3 (n_a^i m_b^i -n_b^i m_a^i).
    \label{eq:Index}
\end{align}
Furthermore, magnetized D-branes carry low-dimensional D-brane charges. 
In particular, the D3-brane tadpole charges are satisfied when
\begin{align}
    \sum_a N_a n_a^1 n_a^2 n_a^3 + \frac{1}{2}N_{\rm flux}=16,
    \label{eq:T6tad1}
\end{align}
where we consider 64 O3-planes. 
As discussed in Ref. \cite{Marchesano:2004xz}, one can realize the supersymmetric brane configurations with the SM-like spectra:
\begin{table}[H]
    \centering
    \begin{tabular}{|c|c|c|c|c|} \hline
        $N_\alpha$ & {\rm Gauge~group} & ($n_\alpha^1, m_\alpha^1$) & ($n_\alpha^2, m_\alpha^2$) & ($n_\alpha^3, m_\alpha^3$)\\ \hline \hline
        $N_a=6$ & $SU(3)_C$ & (1,0) & ($g, 1$) & ($g, -1$)\\ \hline
         $N_b=2$ & $USp(2)_L$ & (0,1) & ($1, 0$) & ($0, -1$)\\ \hline
        $N_c=2$ & $USp(2)_R$ & (0,1) & ($0, -1$) & ($1,0$)\\ \hline
        $N_d=2$ & $U(1)_d$ & (1,0) & ($g, 1$) & ($g, -1$)\\ \hline
    \end{tabular}
\end{table}
Here, the magnetic flux $g$ determines the index of quarks and leptons $I_{ab}=I_{ac}=g$.\footnote{Here, we count the index of both the $\mathbb{Z}_2$-even and -odd modes.} 
From the tadpole cancellation conditions (\ref{eq:T6tad1}), such an index is bounded from above by the orientifold contribution:
\begin{align}
    8g^2 = 16 - \frac{N_{\rm flux}}{2} \leq 16,
\end{align}
i.e., $|I_{ab}|\leq \sqrt{2}$, 
where the three-form fluxes are supposed to be the imaginary self-dual fluxes. 

\medskip

So far, we have not considered the existence of a discrete $B$-field, but such a half-integer flux changes the expression of the tadpole cancellation conditions. 
When we turn on $B$-field on the third torus, the D3-brane tadpole cancellation condition is modified by introducing $\tilde{n}_a^3 = n_a^3 + \frac{1}{2}m_a^3$:
\begin{align}
    \sum_a N_a n_a^1n_a^2\tilde{n}_a^3 +\frac{1}{2}N_{\rm flux} = 8.
    \label{eq:T6tad2}
\end{align}
Following Ref. \cite{Kai:2023ivp}, one can realize the supersymmetric brane configurations with the Standard Model-like spectra:
\begin{table}[H]
    \centering
    \begin{tabular}{|c|c|c|c|c|} \hline
        $N_\alpha$ & {\rm Gauge~group} & ($n_\alpha^1, m_\alpha^1$) & ($n_\alpha^2, m_\alpha^2$) & ($\tilde{n}_\alpha^3, m_\alpha^3$)\\ \hline \hline
        $N_a=8$ & $U(4)_C$ & ($0,-1$) & ($1, 1$) & ($1/2, 1$)\\ \hline
         $N_b=4$ & $U(2)_L$ & ($g,1$) & ($1, 0$) & ($1/2, -1$)\\ \hline
        $N_c=4$ & $U(2)_R$ & ($g,-1$) & ($0, 1$) & ($1/2, -1$)\\ \hline
    \end{tabular}
\end{table}
Here, the magnetic flux $g$ determines the index of quarks and leptons $I_{ab}=I_{ca}=g$. 
From the tadpole cancellation conditions (\ref{eq:T6tad2}), such an index is bounded from above by the orientifold contribution:
\begin{align}
    2g = 8 - \frac{N_{\rm flux}}{2} \leq 8,
\end{align}
i.e., $|I_{ab}|\leq 4$, 
where the three-form fluxes are supposed to be the imaginary self-dual fluxes. 

\medskip

We restrict ourselves to evaluating the explicit bound in concrete toroidal orientifolds, but it is interesting to explore the distribution of the chiral index of matter fields on magnetized D-branes wrapping holomorphic cycles in Calabi-Yau threefolds. 
Since the upper bound on the Atiyah-Singer index will depend on what extent thee-form fluxes $N_{\rm flux}$ distribute inside the Type IIB/F-theory landscape, 
we will leave a comprehensive study about the distribution of the chiral index in the future.

\section{Conclusions}
\label{sec:con}

We studied the index of chiral fermions in the effective action of 
$E_8\times E_8^\prime$ and $SO(32)$ heterotic line bundle models, and magnetized D7-branes 
in Type IIB/F-theory compactifications. 
We derived an analytical expression for the upper bound of the chiral index, which depends on the topological quantities of CY manifolds. 
It results in a finite number of consistent line bundle models subject to phenomenological constraints on favorable CYs. 
We focused on a limited class of consistent string models, but it is interesting to extend our analysis into other corners of the string landscape in the future.

\medskip

In $E_8\times E_8^\prime$ and $SO(32)$ heterotic line bundle models, 
we evaluated the upper bound on complete intersection CY threefolds. 
Since the CY volume is bounded from above by ${\cal V}\leq 25$ to realize the observed value of 4D gauge couplings, 
semi-realistic models are realized on some of the complete intersection CY threefolds. 
On quotient CY threefolds, the index of chiral fermions decreases 
due to the freely-acting discrete symmetry group. 
It turned out that the chiral index satisfies $N_{\rm gen}\leq 100$ on quotient CY threefolds we analyzed. 
We also commented on the favorable CY threefolds with the large number of K\"ahler moduli in the Kreuzer-Skarke database. 
Note that for $SO(32)$ heterotic string theory, we found the bound on the chiral index of Higgs doublets and vector-like quarks in addition to quarks and leptons. 
Our finding bound on the chiral index is correlated with the second Chern number of the tangent bundle of CY threefolds. Such a mutual relation between the chiral index and the second Chern number was also pointed out in the context of a machine learning study \cite{Otsuka:2020nsk}, and it would be interesting to figure out a physical background of this suggestive behavior.

\medskip

In Type IIB/F-theory compactifications, we focused on the index of chiral fermions arising in the intersection between 
two stacks of magnetized D7-branes. 
Since the world-volume fluxes on D7-branes play a role not only in determining the chiral index but also in 
inducing the D3-brane charges, the D3-brane tadpole cancellation condition restricts the maximal value of flux quanta. 
It turned out that the chiral index on matter divisors or matter curves is restricted by the number of D7-brane and 
topological quantities of CY manifolds. 
For concreteness, we analyzed the generation number of quarks and leptons on $T^6/(\mathbb{Z}_2\times \mathbb{Z}_2^\prime)$ orientifolds. The index is indeed bounded by the orientifold contributions. 
Thus, the small tadpole charge leads to a few generations of chiral fermions.

\acknowledgments

The authors would like to thank Andrei Constantin for useful discussions. 
This work was supported in part by Kyushu University’s Innovator Fellowship Program (S.N.), JSPS KAKENHI Grant Numbers JP20K14477 (H.O.), JP 21J20739 (M.T.), JP22J12877 (K.I.) and JP23H04512 (H.O).

\bibliography{referencesv2}{}
\bibliographystyle{JHEP} 

\end{document}